\documentclass[runningheads]{llncs}
\usepackage[T1]{fontenc}
\usepackage{xcolor} 
\usepackage{float}
\usepackage[hidelinks]{hyperref}

\usepackage{graphicx}
\usepackage{multirow}
\usepackage{amssymb}
\usepackage{booktabs}
\usepackage[numbers,sort]{natbib}
\usepackage[usestackEOL]{stackengine}

\begin{document}
\title{Automated Spinal MRI Labelling from Reports Using a Large Language Model}

\titlerunning{Automated Spinal MRI Labelling}

\author{Robin Y. Park, Rhydian Windsor, Amir Jamaludin, and Andrew Zisserman}

\authorrunning{R. Y. Park et al.}
%
\institute{Visual Geometry Group, Department of Engineering Science,\\University of Oxford, Oxford, UK\\
\email{robinpark@robots.ox.ac.uk}}
\maketitle              

\begin{abstract}
We propose a general pipeline to automate the extraction of labels from radiology reports using large language models, which we validate on spinal MRI reports. The efficacy of our labelling method is measured on five distinct conditions: spinal cancer, stenosis, spondylolisthesis, cauda equina compression and herniation. Using open-source models, our method equals or surpasses GPT-4 on a held-out set of reports. Furthermore, we show that the extracted labels can be used to train imaging models to classify the identified conditions in the accompanying MR scans. All classifiers trained using automated labels achieve comparable performance to models trained using scans manually annotated by clinicians\footnote{Code can be found at \url{https://github.com/robinyjpark/AutoLabelClassifier}.}.

\end{abstract}

\section{Introduction}
Labelling medical image datasets can be time-consuming and requires expert annotators, whose time is limited and expensive. This is compounded by a large number of medical conditions that can occur in any given image and often large inter-reader variability leading to noisy labels. This means researchers applying deep learning to medical imaging problems usually settle for small-scale datasets compared to other areas in machine learning, spend large amounts of funding on data collection, or restrict themselves to a few publicly-available datasets covering only a few conditions and modalities. A possible solution to this problem is to extract labels directly from radiological reports -- free-text descriptions written by radiologists describing the findings of an imaging investigation. These reports can be downloaded in bulk from a hospital's electronic database; if the extraction can be automated, it would significantly reduce the annotation bottleneck and unlock much larger training datasets for solving medical imaging problems. 

However, automated extraction of structured information from clinical reports is not a new idea and has proved to be challenging~\cite{thirunavukarasu2023large}, due to large variability in reporting styles, heavy use of domain-specific vocabulary and frequently assumed knowledge (e.g. a report describing an investigation into the presence of metastasis implies the existence of a primary cancer). We propose a general method for extracting structured labels for vision models from clinical reports. This is achieved by adapting general-purpose large language models (LLMs) by asking the model to summarise the report with a target condition in mind and produce a binary label based on the summary. Crucially, to obtain labels for a new medical condition, all that is required is the class name and a definition but no further fine-tuning.

To test the efficacy of this method, we apply it to spinal magnetic resonance imaging (MRI) radiological reports, aiming to label spinal cancer, stenosis, spondylolisthesis, cauda equina compression and herniation. Adapting Zephyr (7B) and Llama3 Instruct (8B), two open-source question-answering LLMs, our method achieves balanced accuracy and F1 scores equal to or exceeding that of GPT-4 for both conditions. We then use the extracted labels to train a vision model to detect cancer, stenosis and spondylolisthesis in spinal MRI. The classification networks trained using the automated labels are able to match the performance of existing classifiers trained using large volumes of expert-annotated images.

\subsection{Related Work}

Rule-based report parsers have shown strong performance in many settings (e.g.\ CheXpert~\cite{irvin2019chexpert}, DeepSpine~\cite{deepspine}). However, developing these parsers requires domain knowledge and cannot be easily adapted to new conditions, which conflicts with our aim of reducing reliance on expert input and manual annotations. 

LLMs have become increasingly accessible to researchers with the emergence of open-source models such as Llama~\cite{touvron2023llama}, Alpaca~\cite{alpaca} and Mistral~\cite{jiang2023mistral}. Methods like instruction fine-tuning~\cite{ouyang2022training} can improve models' alignment with question-answering tasks, and low rank adaptation (LoRA)~\cite{hu2021lora} allows training large models with limited compute. Many existing models have been further pre-trained on biomedical data to increase the base model's familiarity with biomedical vocabulary and syntax~\cite{gu2021domain, li2024llava}. However, these are often trained using clinical abstracts and papers, which have a different form to radiological reports~\cite{lee2020biobert,luo2022biogpt, li2024llava} or use reports from a single modality, e.g.\ Chest X-rays~\cite{boeckingbiovil,bannur2023learning}. RadBERT~\cite{yan2022radbert} is an example of a model pre-trained using a large corpus of diverse radiology reports; however, this is a 110M parameter BERT-based model and is thus unlikely to adapt to new tasks as well as much larger publicly-available general models. 

Accordingly, as more powerful LLMs continue to emerge, there has been a growing focus on the development of generalisable methods that do not require specialised training on the task at hand. Steering GPT-4 by asking the model to come up with its own chain-of-thought has achieved state-of-the-art results, demonstrating significant gains in accuracy on medical question-answering over specialist fine-tuned models like Med-PaLM 2~\cite{nori2023can}. While GPT-4 achieves excellent performance across many specialist domains, costs can be high when processing large datasets. Furthermore, closed-source LLMs accessible through APIs like GPT-4 require data to be uploaded to remote servers for processing, which poses a privacy risk, especially for medical reports containing potentially sensitive information. Finally, since weights are not publicly available, the model cannot be customised or incorporated into vision-language models downstream. 

\section{Extracting Structured Labels from Radiological Reports Using a Large Language Model}

\begin{figure}[h]
\centering
\includegraphics[width=\textwidth]{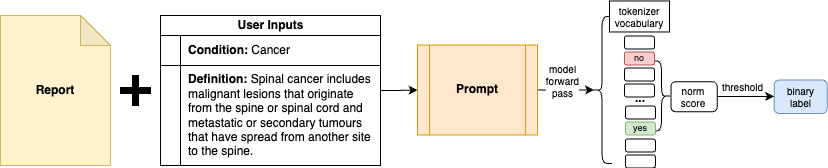}
\caption{\label{label_process} \textbf{Radiological report labelling pipeline:} The prompt step formats the user inputs as shown in Figure \ref{prompts} to summarise the report based on the target condition. Based on this summary, we extract the binary label using the normalised scores from a chosen set of two unique tokens (``yes'' and ``no'') in the vocabulary.}
\end{figure}

Figure~\ref{label_process} shows an overview of our method to extract labels from clinical reports. The user specifies the condition for classification along with a definition to reduce ambiguity. These are input into a general-purpose prompting template (see Figure~\ref{prompts}). Our method has two steps: (1) asking the model to generate a summary of the report based on the target condition, and (2) using the summary to assign a binary label. We also perform self-supervised fine-tuning to familiarise the model with the summary generation task, which is described in Section~\ref{sec:summary-finetuning}.

\subsection{Model Prompting}

\begin{figure}
\centering
\includegraphics[width=.9\textwidth]{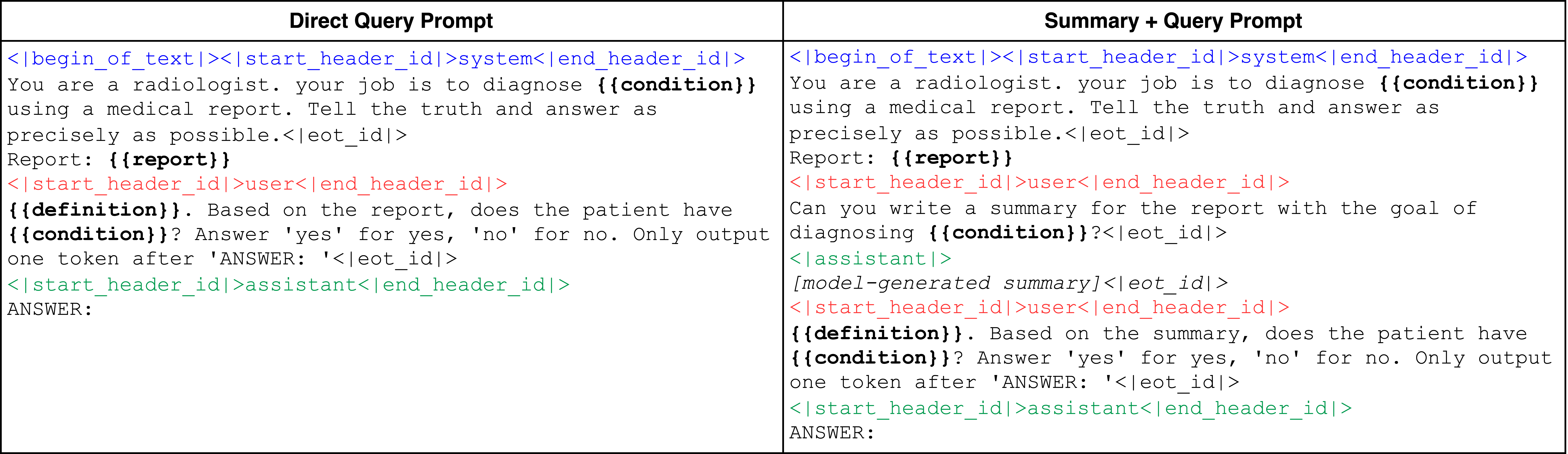}
\caption{\label{prompts} \textbf{Model prompting strategies:} The direct query method (left) asks the model to extract the label based on the report. The summary and query method (right) asks the model to generate a summary focused on the condition, which it uses as additional input to annotate the report. Words in bold indicate user inputs to be modified.}
\end{figure}

We tested our prompting method using Zephyr-7B and Llama-8B Instruct, two instruction-fine-tuned language models that can support a wide range of use cases. To prompt the models, we provide a definition of the condition of interest. We tested two methods of prompting: (1) asking the model directly whether the patient has the condition based on the report, and (2) requesting that the model generates a summary of the report based on the condition of interest and decide whether the patient has the condition based on that summary. We formulated the prompt to extract binary labels on a given condition as seen in Figure~\ref{prompts}. To generate these labels, we used the softmaxed logits of two unique tokens in the tokenizer's vocabulary: ``yes'' and ``no''; see Figure~\ref{label_process}. Using the token scores ensures that the model will produce a binary answer to every question.

\subsection{Domain Adaptation By Summary Fine-Tuning}
\label{sec:summary-finetuning}

Radiology reports often include summary sections, which give overviews of the findings. These summaries typically include the most pertinent information, including potential diagnoses and descriptions of disease progression~\cite{peng2018negbio,irvin2019chexpert}. Thus, the summary is especially relevant to consider when extracting clinical labels. 

To ensure that the model would generate clinically relevant summaries, we fine-tuned the linear layers of Zephyr using LoRA~\cite{hu2021lora} to perform next token prediction on spinal reports from a local hospital system. To do this, we identified the summary sections of 56,924 reports using regular expression matches for the following words (case-insensitive): conclusion, impression, findings, and summary (of the 124,771 reports available in the dataset, only 56,924 reports had matches for one of these words). While we fed the whole report to the model as context, the next-token-loss was computed only using the summary section of the report.

\section{Supervised Learning Using Automated Labels}

\begin{figure}[h]
\centering
\includegraphics[width=.9\textwidth]{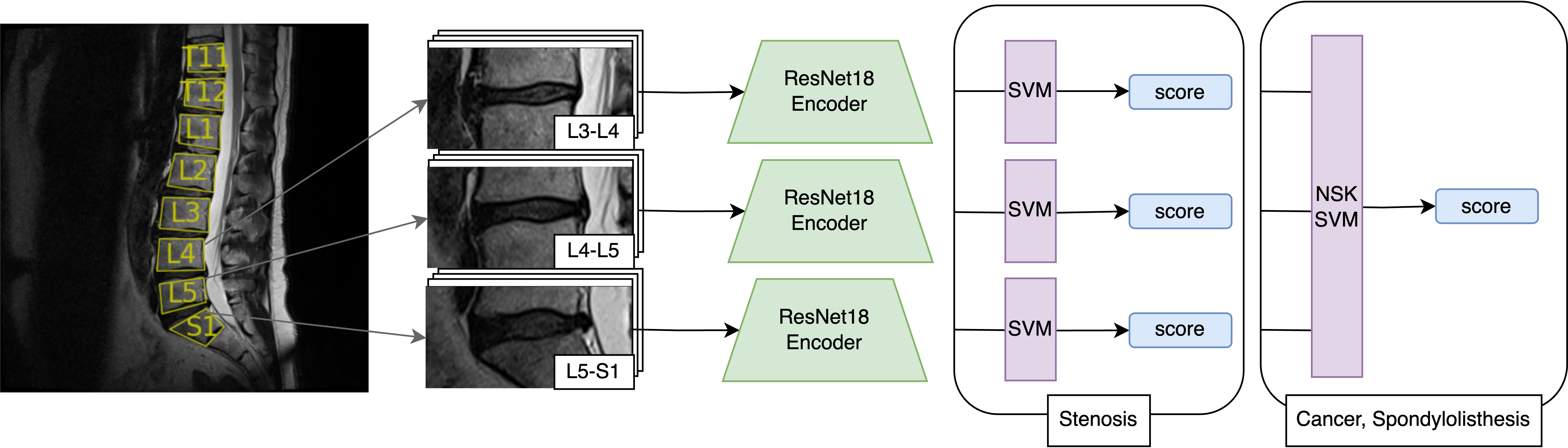}
\caption{\label{img_cls_pipe} \textbf{MRI classification network:} SpineNetV2 is used to detect IVDs. Each IVD is encoded using ResNet18. For stenosis, we use a SVM to get a score per IVD. For cancer and spondylolisthesis, we aggregate IVD encodings and use NSK-SVM to get a score per scan.}
\end{figure}

To demonstrate that the labels automatically generated by our pipeline can be used to train a vision model, we used the generated labels to train a classifier to detect cancer, stenosis and spondylolisthesis in the MR scans paired to the reports. We were not able to train classifiers for cauda equina compression and herniation due to insufficient numbers of positive cases in our dataset.

SpineNetV2~\cite{windsor2024automated} was used to automatically detect the vertebrae and extract the intervertebral discs (IVDs) in the T2 sagittal scans. Each IVD is of dimension slice x height x width ($9 \times 112 \times 224$). We reshaped each slice to $224 \times 224$ to input them into a modified ResNet18 (pre-trained on ImageNet) without the fully connected layer to get slice-level embeddings. These were averaged across slices to get the volume-level representations.  The embeddings were used as features to train support vector machines (SVM) with a linear kernel to perform binary classification for our conditions of interest. For stenosis, we could generate level-specific labels, so we detected stenosis at each IVD level. For cancer and spondylolisthesis, the labels are provided at the spine-level but image samples are individual IVDs, so we used multiple-instance learning (MIL) using the Normalised Set Kernel method (NSK-SVM) to get the average representation across IVDs in a spinal scan~\cite{gartner2002multi}. Section \ref{sec:datasets} provides more information on the granularities of the report-generated labels. Figure~\ref{img_cls_pipe} provides an illustration of the full pipeline. 

\section{Datasets}\label{sec:datasets}

We use three datasets: \textbf{CancerData}, \textbf{LumbarData}, and \textbf{GeneralReports}. The first two contain report-image pairs while the last contains reports only.

\textbf{Label Extraction:} A subset of \textbf{CancerData} and \textbf{LumbarData} was manually labelled for prompt engineering experiments and testing our labelling pipeline. \textbf{CancerData} were provided by National Consortium of Intelligent Medical Imaging (NCIMI) and include reports and the associated MRIs of confirmed or potential cancer cases from six different NHS Trusts across the UK. \textbf{LumbarData} were collected as part of the Oxford Secondary Care Lumbar MRI Cohorts (OSCLMRIC) study and includes clinical MRI studies and reports of patients with lower back pain, sourced from a local hospital system. Samples of \textbf{LumbarData} were used to label positive and control cases for stenosis, spondylolisthesis, cauda equina and herniation. Each manually annotated dataset were a randomly chosen subset of the full source data, which we split into a calibration set (to develop our prompting strategy and determine a model calibration threshold) and a test set, using stratified sampling.  Manual labels for \textbf{Cancer} were provided by a radiology registrar. Manual labels for \textbf{Stenosis} were provided by an orthopaedic surgeon. \textbf{Spondylolisthesis}, \textbf{Cauda Equina} compression and \textbf{Herniation} were labelled by the author. \textbf{GeneralReports} were also provided by OSCLMRIC and contain 56,924 unpaired spinal MRI reports with diverse conditions and control examples; it was used to fine-tune the model on the causal language modeling task to generate the summary. See Table~\ref{labdata_tab} for data splits for each task.

The reports were preprocessed based on use case. Spinal reports are often split into sections (see Figure~\ref{report_sections} in Appendix for an example). Since spinal cancers are usually metastases from a primary site elsewhere, clinical histories in \textbf{CancerData} frequently mention the presence of cancer in a non-imaged site and includes query words (i.e.\ ``?cancer'' to check for spinal cancer in the image). To correctly identify spinal cancer (i.e.\ the condition that would be visible in the paired image), we excluded clinical history when inputting reports from \textbf{CancerData} into the pipeline. As stenosis, spondylolisthesis, cauda equina compression and herniation are specific to the spine, we did not perform any additional preprocessing on \textbf{LumbarData}. For the summary fine-tuning task, we identified the summary sections of the reports such that the rest of the report could be masked for loss computation (see Section~\ref{sec:summary-finetuning}).

\begin{table}[h]
\footnotesize
\centering
\setlength{\tabcolsep}{5pt}
\begin{tabular}{ccccc}
\hline
\textbf{Dataset} & \textbf{Split} & \textbf{Patients} & \textbf{Studies} & \textbf{Positive Cases} \\ \hline
\multirow{2}{*}{\textbf{Cancer}}   & Calibration (Labels) & 140    & 140    & 67 (47.9\%)  \\
                              & Test (Labels)     & 145    & 145    &  67 (46.2\%)  \\ \hline
\multirow{2}{*}{\textbf{Stenosis}} & Calibration (Labels) & 68     & 68     & 41 (60.3\%)  \\
                              & Test (Labels)     & 68     & 68     & 45 (66.2\%)  \\ \hline
\multirow{2}{*}{\textbf{Spondylolisthesis}} & Calibration (Labels) & 76     & 76     & 33 (43.4\%)  \\
                              & Test (Labels)    & 77     & 77     & 38 (49.4\%)  \\ \hline   
\multirow{2}{*}{\textbf{Cauda Equina}} & Calibration (Labels) & 94     & 94     & 18 (19.1\%)  \\
                              & Test (Labels)    & 94     & 94     & 17 (18.1\%)  \\ \hline
\multirow{2}{*}{\textbf{Herniation}} & Calibration (Labels) & 70     & 70     & 28 (40.0\%)  \\
                              & Test (Labels)    & 70     & 70     & 28 (40.0\%)  \\ \hline  \hline  
\textbf{GeneralReports}                & Fine-tuning          & 44,952 & 56,924 & -   \\ \hline       
\end{tabular}
\caption{\label{labdata_tab} \textbf{Labelling datasets} used to prompt engineer and fine-tune the model for the labelling pipeline. \textbf{Cancer} splits are subsets of \textbf{CancerData}. \textbf{Stenosis}, \textbf{Spondylolisthesis}, \textbf{Cauda Equina} and \textbf{Herniation} are subsets of \textbf{LumbarData}.} 
\end{table}

\begin{table}[h]
\centering
\footnotesize
\setlength{\tabcolsep}{2.2pt}
\begin{tabular}{ccccccc}
\hline
\multicolumn{1}{c}{\textbf{}} &
  \multicolumn{1}{c}{\textbf{}} &
  \multicolumn{1}{c}{\textbf{}} &
  \multicolumn{1}{c}{\textbf{}} &
  \multicolumn{3}{c}{\textbf{IVDs}} \\ \cline{5-7} 
\textbf{Condition} &
  \textbf{Split} &
  \textbf{Patients} &
  \textbf{Studies} &
  \textbf{Total} &
  \textbf{Positives} &
  \textbf{Negatives} \\ \hline
\multirow{3}{*}{Cancer} & Train (Scans)   & 1,223 & 1,256 & 14,167 & 9,012 & 5,155~ \\
                                       & Validation (Scans) & 324  & 327  & 3,612  & 2,149  & 1,463~  \\
                                       & Test (Scans)    & 450  & 451  & 4,918 & 3,120  & 1,798~  \\ \hline 
\multirow{3}{*}{Stenosis} & Train (Scans)   & 1,375 & 1,946 & 5,827 & 2,977 & 2,850~ \\
                                       & Validation (Scans) & 153  & 217  & 649  & 337  & 312~  \\
                                       & Test (Scans)    & 117  & 123  & 368  & 139  & 229~  \\ \hline 
\multirow{3}{*}{\shortstack{Spondylo-\\listhesis}} & Train (Scans)   & 1,357 & 1,939 & 5,758 & 725 & 5,033~ \\
                                       & Validation (Scans) & 152  & 200  & 645 & 78  & 567~  \\
                                       & Test (Scans)    & 146  & 151  &  453  & 213 & 240~  \\ \hline \\
\end{tabular}
\caption{\label{scan_tab} \textbf{Summary of splits for MRI classification}. The cancer splits are subsets of \textbf{CancerData}. The stenosis and spondylolisthesis splits are subsets of \textbf{LumbarData}. Positive/negative labels on the training and validation sets were assigned using our labelling method described in section~\ref{sec:results-label-acc} whereas test set labels were manually annotated as described in Section~\ref{sec:datasets}. Test (Scans) in each dataset consists of studies from both Calibration and Test sets in Table~\ref{labdata_tab} where the IVDs were successfully extracted.} 
\end{table}
 
\textbf{MRI Classification:} For the MRI classification tasks, we used T2 sagittal images in a valid report-image pair in \textbf{CancerData} and \textbf{LumbarData}. \textbf{LumbarData} had only one sequence per study. \textbf{CancerData} could have multiple; in these cases, we chose the latest whole spine sequence in the study.

For stenosis, we generated labels for each IVD using our pipeline. This was feasible for stenosis as (1) reports almost always contained information about the level at which stenosis is present, and (2) report-level expert annotations were level-specific (for lumbar IVDs T12-S1). We only used the last three lumbar spine IVDs  (L3-L4, L4-L5 and L5-S1) as stenosis examples above L3 are rare. It was less common for the cancer reports to list specific levels at which cancer is present, especially if present at multiple levels, which is common for both metastases and myeloma. As a result, we follow the approach in~\cite{Windsor22a} and employ multiple instance learning, treating the entire spine as a bag which is labelled as positive if any of the vertebrae show cancer, or negative otherwise. We included IVDs from cervical to lumbar (18 IVDs C7-S1). Table~\ref{scan_tab} summarises the splits.

\section{Results} 

\begin{figure}[H]
\centering
\includegraphics[width=\linewidth]{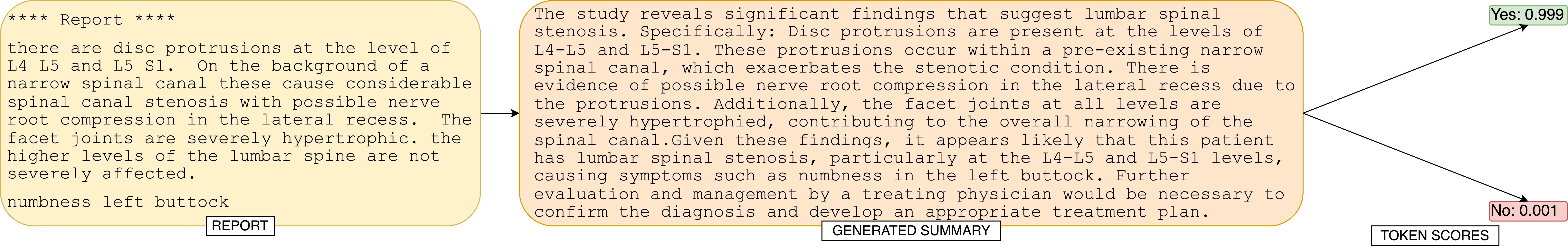}
\caption{\label{summ_result} Real example report from \textbf{LumbarData} with the summary and scores generated for \textbf{Stenosis} using our pipeline. Any dates, names or location information were removed from the report.}
\end{figure}

\subsection{Assessment of Labelling Accuracy}\label{sec:results-label-acc}

We evaluated the model's performance on label extraction by measuring area under the receiver-operator curve (AUROC) and equal error rate (EER) using the normalised scores from the model. To compute balanced accuracy and F1 score, we used the EER threshold from the calibration set.

\begin{table}[H] 
\setlength{\tabcolsep}{3.1pt}
\centering
\footnotesize
{\begin{tabular}{clcccccc}
\hline 
 \textbf{Condition} & \textbf{\shortstack{Prompt \\ Method}} & \textbf{\shortstack{Summary \\ Model}} & \textbf{\shortstack{Query \\ Model}} & \textbf{\shortstack{Bal. \\ Acc.}} & \textbf{EER} & \textbf{\shortstack{AU \\ ROC}} & \textbf{F1} \\ \hline
\multirow{5}{*}{Cancer} & Direct Query & - & GPT-4 & \textbf{1.000} & -  & -  & \textbf{1.000} \\
& Direct Query & - & Zephyr & 0.993 & \textbf{0.000} & \textbf{1.000} & 0.992 \\
& Direct Query & - & Llama3 & 0.978 & 0.014 & 0.999 & 0.977 \\
& Summary-Query & Zephyr & Zephyr & 0.985 & \textbf{0.000} & \textbf{1.000} & 0.985 \\
& Summary-Query & Z-SFT & Zephyr & 0.994 & \textbf{0.000} & \textbf{1.000} & 0.993 \\ 
& Summary-Query & Llama3 & Llama3 & \textbf{1.000} & \textbf{0.000} & \textbf{1.000} & \textbf{1.000} \\ \hline
\multirow{5}{*}{Stenosis} & Direct Query  & - & GPT-4  & \textbf{0.951} & -     & -     & \textbf{0.949} \\
& Direct Query  & -  & Zephyr & 0.945 & \textbf{0.037} & 0.981 & 0.950 \\
& Direct Query & - & Llama3 & \textbf{0.963} & \textbf{0.000} & \textbf{0.987} & \textbf{0.962} \\
& Summary-Query & Zephyr & Zephyr & 0.945 & 0.111 & 0.985 & 0.950 \\
& Summary-Query & Z-SFT & Zephyr & 0.933 & \textbf{0.037} & 0.943 & 0.937  \\ 
& Summary-Query & Llama3 & Llama3 & 0.945 & \textbf{0.037} & \textbf{0.995} & 0.950  \\ \hline
\multirow{5}{*}{\shortstack{Spondylo-\\listhesis}} & Direct Query  & - & GPT-4  & \textbf{0.974} & -     & -     & 0.973 \\
& Direct Query  & -  & Zephyr & \textbf{0.974} & \textbf{0.026} & \textbf{0.996} & \textbf{0.974} \\
& Direct Query & - & Llama3 & \textbf{0.974} & \textbf{0.026} & 0.993 & \textbf{0.974} \\
& Summary-Query & Zephyr & Zephyr & 0.942 & \textbf{0.023} & 0.985 & 0.938 \\
& Summary-Query & Z-SFT & Zephyr & 0.948 & \textbf{0.026} & 0.984 & 0.946 \\ 
& Summary-Query & Llama3 & Llama3 & \textbf{0.974} & \textbf{0.026} & \textbf{0.994} & \textbf{0.974}\\ \hline
\multirow{5}{*}{\shortstack{Cauda Equina\\Compression}} & Direct Query  & - & GPT-4  & \textbf{1.000} & -     & -     & \textbf{1.000} \\
& Direct Query  & -  & Zephyr & 0.912 & \textbf{0.013} & 0.972 & 0.903 \\
& Direct Query & - & Llama3 & 0.941 & 0.026 & \textbf{0.998} & 0.938 \\
& Summary-Query & Zephyr & Zephyr & 0.971 & 0.039 & \textbf{0.998} & 0.970 \\
& Summary-Query & Z-SFT & Zephyr & 0.882 & \textbf{0.013} & 0.988 & 0.867  \\ 
& Summary-Query & Llama3 & Llama3 & \textbf{1.000} & \textbf{0.000} & \textbf{1.000} & \textbf{1.000}  \\ \hline
\multirow{5}{*}{Herniation} & Direct Query  & - & GPT-4  & \textbf{0.935} & -     & -     & \textbf{0.915} \\
& Direct Query  & -  & Zephyr & 0.869 & 0.119 & 0.947 & 0.842 \\
& Direct Query & - & Llama3 & 0.911 & \textbf{0.071} & \textbf{0.980} & 0.893 \\
& Summary-Query & Zephyr & Zephyr & 0.774 & 0.310 & 0.815 & 0.738 \\
& Summary-Query & Z-SFT & Zephyr & 0.768 & 0.214 & 0.871 & 0.730  \\ 
& Summary-Query & Llama3 & Llama3 & \textbf{0.946} & \textbf{0.071} & \textbf{0.990} & \textbf{0.931} \\ \hline
\end{tabular}}
\caption{\label{label_tab} \textbf{Scan-level report labelling performance} on cancer, stenosis, spondylolisthesis, cauda equina compression and herniation test sets (see Table~\ref{labdata_tab}). Normalised scores are used to compute EER and AUROC. Calibrated binary labels are used to compute balanced accuracy and F1. \textit{Zephyr} refers to the base model without fine-tuning, whereas \textit{Z-SFT} indicates the summary fine-tuned model (see \S\ref{sec:summary-finetuning}).  The top two best values per metric are bolded.} 
\end{table}

Table~\ref{label_tab} shows the results of labelling each condition on the respective test sets. Across all tasks, our best methods exceed 0.94 balanced accuracy and match or outperform a direct query using GPT-4, prompted using the same format and input as the direct query strategy in Figure~\ref{prompts} but with the tags (e.g.\ user, system, assistant) formatted for GPT-4. For cancer, the summary-query method using Llama3 achieves perfect balanced accuracy, AUROC and F1. For stenosis, the summary-query method using Llama3 achieves the highest AUROC; the direct query method using Llama3 achieves higher balanced accuracy and F1. For cauda equina, both direct query and summary-query using Llama3 achieve best performance, matching GPT-4. Herniation is the hardest task for GPT-4 and Zephyr; Llama3 summary-query achieves the best balanced accuracy, AUROC and F1. 

\begin{table}[h] 
\centering
\small
\setlength{\tabcolsep}{5pt}
{\begin{tabular}{ccccc}
\hline
\textbf{IVD Level} & \textbf{Bal. Acc.} & \textbf{EER} & \textbf{AUROC} & \textbf{F1} \\ \hline
\textbf{L3-L4}     & 0.896              & 0.073        & 0.968          & 0.815       \\
\textbf{L4-L5}     & 0.908              & 0.108        & 0.978          & 0.900       \\
\textbf{L5-S1}     & 0.855              & 0.150        & 0.945          & 0.830       \\ \hline
\end{tabular}}
\caption{\label{ivd_test} \textbf{IVD-level report labelling performance} on expert-annotated subset of \textbf{LumbarData} for \textbf{Stenosis} (n=68 patients, 204 IVDs). Normalised scores are used to compute EER and AUROC. Calibrated binary labels are used to calculate balanced accuracy and F1.}
\end{table}

Since it achieved the highest or second highest AUROC across all tasks, the summary-query prompting method with Llama3 for summary generation and Q-A was used as the general method to extract binary labels for the MRI classification tasks. Figure~\ref{summ_result} shows an example of the summary and token scores generated from our pipeline. For inference across the full datasets, we extracted scan-level labels for cancer and IVD-level labels for stenosis, as described in section~\ref{sec:datasets}. Table~\ref{ivd_test} shows the performance of our labelling pipeline for stenosis at each IVD. 

\subsection{Assessment of MRI Classification Accuracy}

\begin{table}[h]
\small
\centering
\setlength{\tabcolsep}{1.1pt}
{\begin{tabular}{ccccccc}
\hline
& \textbf{Condition} & \textbf{Model}  & \textbf{Bal. Acc.} & \textbf{EER} & \textbf{AUROC} & \textbf{F1} \\ \hline
(1) & \textbf{Cancer}   & ResNet18+NSK-SVM &   0.763       &           0.215        &   0.851             &   0.773          \\ \hline
(2) &\textbf{Stenosis}  & SpineNetV2~\cite{windsor2024automated} & 0.679             & -              & -          & 0.574       \\ 
(3) & \textbf{Stenosis}  & SpineNetV2~\cite{windsor2024automated}+SVM & 0.780             & 0.218              & 0.857          & 0.727       \\ 
(4) & \textbf{Stenosis}  & ResNet18+SVM & 0.775             & 0.227              & 0.836          & 0.722       \\ \hline
(5) &\textbf{Spondylolisthesis}  & SpineNetV2~\cite{windsor2024automated} & 0.861             & -              & -          & 0.853      \\ 
(6) & \textbf{Spondylolisthesis}  & SpineNetV2~\cite{windsor2024automated}+NSK-SVM & 0.858             & 0.138              & 0.910         & 0.855       \\ 
(7) & \textbf{Spondylolisthesis}  & ResNet18+NSK-SVM & 0.780             & 0.250              & 0.855          & 0.785       \\ 
\hline
\end{tabular}}
\caption{\label{cls_acc} \textbf{MRI Classifier Performance} on \textbf{CancerData} and \textbf{LumbarData} test sets. Calibrated binary labels are used to calculate balanced accuracy and F1. Only rows (1), (4) and (7) report results of models fully trained using our labels. (2) and (5) perform inference using SpineNetV2, which is trained using human annotations, and (3) and (6) use SpineNetV2 to extract encodings and train an SVM using our report-generated labels. We treat (2)-(3) as baselines for our stenosis classification method (4) and (5)-(6) as baselines for our spondylolisthesis classification method (7).}
\end{table} 

We evaluated the model's performance on image classification by measuring area under the receiver-operator curve (AUROC) and equal error rate (EER) using the normalised scores from the model. We used the EER threshold from the validation set to derive binary labels from the scores on the test set, which were used to compute balanced accuracy and F1 score.

\textbf{Stenosis:} SpineNetV2~\cite{windsor2024automated} has an existing grading model trained to classify three kinds of stenosis (central canal, foraminal right and left). We compare our method (ResNet18+SVM) to (1) an aggregated stenosis label from SpineNetV2 and (2) SVM using SpineNetV2 encodings (SpineNetV2+SVM). Our stenosis classifier outperforms aggregated SpineNetV2 stenosis scores and achieves a similar AUROC, balanced accuracy and F1 score after calibration for either choice of encoding with the SVM trained on automated labels (see Table~\ref{cls_acc}). 

\textbf{Spondylolisthesis:} Similarly to stenosis, SpineNetV2~\cite{windsor2024automated} has an existing grading model trained to generate a binary label for spondylolisthesis. We compare our method (ResNet18+NSK-SVM) to (1) the label from SpineNetV2 and (2) NSK-SVM using SpineNetV2 encodings (SpineNetV2+NSK-SVM). While the SpineNetV2 labels and NSK-SVM trained on SpineNetV2 encodings exceed our model's performance, our model achieves good performance (AUROC: 0.855) comparable to the cancer and stenosis classifiers.

\textbf{Cancer:} Considering all IVD-level embeddings in aggregate, our classifier achieves a balanced accuracy of 0.763 in classifying cancer at the scan-level. SpineNetV2 does not classify cancer, so we are unable to provide a comparable baseline on the same dataset. A deep learning model to detect cancer in CT images achieved an F1 score of 0.72~\cite{motohashi2024new}, which we surpass (0.773).

\subsection{Discussion and Limitations}

There are several training and prompting methods we tested but did not include in our final methodology. While adding few-shot examples of desired Q-A pairs when prompting LLMs is a common method to improve performance~\cite{brown2020language,nori2023can}, we found results to be highly sensitive based on the reports selected as examples. Furthermore, adding longer example medical reports often exceeds the models' context window (512 tokens for Zephyr, 8,000 tokens for Llama3). Interestingly, we found that fine-tuning Zephyr on report summaries was only beneficial for the cancer labelling task (Table~\ref{label_tab}). We currently force the pipeline to output a positive or negative label without uncertainty. In future work, we plan to work directly with probabilities from our locally-run LLM to predict uncertainty.

\subsection{Recommendations}

Our method can be adapted for use with other open-source LLMs, clinical conditions and reports accompanying other image modalities (e.g. chest X-rays).

For smaller LLMs and clinical conditions that the LLM may not have encountered in its training corpus, we recommend domain adaptation by summary fine-tuning to familiarise the model with the vocabulary and syntax of radiology reports and generate clinically relevant summaries for the target condition (see Section \ref{sec:summary-finetuning}).

Stronger LLMs may not need additional fine-tuning and achieve good results even with the direct query method. For well-defined (``easier'') clinical conditions, the default 0.5 threshold may be sufficient to achieve a good classification. For conditions that are rarely found in the dataset (<20\% of the source data), we recommend using a manually labelled subset of the data to compute the equal error rate and adjust the acceptance threshold to account for the data imbalance.

Additionally, for conditions that can be conflated with related conditions (e.g. herniation and protrusion), it may be necessary to define the negative case as part of the condition's definition (i.e. ``Herniation is a more severe condition than disc protrusion or bulging.''). 

\section{Conclusion}

We propose a general method that can be adapted to extract labels from radiological reports without additional model training. We show that this surpasses a strong GPT-4 baseline when applied to spinal MR reports, with the additional advantages that: (1) we use a locally-run open-source model that is privacy-preserving and cheap, and (2) we have direct access to the raw token-level scores, which can be used for model calibration to produce more accurate binary labels. We also demonstrate that the extracted labels can be used to train a classifiers with similar performance to models trained with expert-annotated scans.

\begin{credits}
{\small}
\subsubsection{\ackname} 
We thank our clinical collaborators for providing annotations and advice: Prof.\ Jeremy Fairbank, Dr.\ Sarim Ather, Prof.\ Iain McCall, Dr.\ Mark Kong (in no particular order). We are also grateful to our funders: EPSRC CDT in Health Data Science (EP/S02428X/1), Cancer Research UK via the EPSRC CDT in Autonomous Intelligent Machines and Systems (EP/S024050/1), EPSRC programme grant Visual AI (EP/T025872/1), and the Oxford Big Data Institute. \textbf{LumbarData} (OSCLMRIC) was collected as part of a Health Research Authority study (IRAS Project ID 207858), while \textbf{CancerData} was collected by NCIMI (Project ID 010A).

\subsubsection{\discintname}
The authors have no competing interests to declare that are relevant to the content of this article.
\end{credits}

\bibliographystyle{splncs04}
\bibliography{Paper-1510}

\begin{thebibliography}{10}
\providecommand{\url}[1]{\texttt{#1}}
\providecommand{\urlprefix}{URL }
\providecommand{\doi}[1]{https://doi.org/#1}

\bibitem{bannur2023learning}
Bannur, S., Hyland, S., Liu, Q., P\'{e}rez-Garc\'{i}a, F., Ilse, M., Castro, D.C., Boecking, B., Sharma, H., Bouzid, K., Thieme, A., Schwaighofer, A., Wetscherek, M., Lungren, M.P., Nori, A., Alvarez-Valle, J., Oktay, O.: Learning to exploit temporal structure for biomedical vision{\textendash}language processing. In: CVPR (2023)

\bibitem{boeckingbiovil}
Boecking, B., Usuyama, N., Bannur, S., Castro, D.C., Schwaighofer, A., Hyland, S., Wetscherek, M., Naumann, T., Nori, A., Alvarez-Valle, J., Poon, H., Oktay, O.: Making the most of text semantics to improve biomedical vision--language processing. In: ECCV (2022)

\bibitem{brown2020language}
Brown, T., Mann, B., Ryder, N., Subbiah, M., Kaplan, J.D., Dhariwal, P., Neelakantan, A., Shyam, P., Sastry, G., Askell, A., et~al.: Language models are few-shot learners. In: NeurIPS (2020)

\bibitem{gartner2002multi}
G{\"a}rtner, T., Flach, P.A., Kowalczyk, A., Smola, A.J.: Multi-instance kernels. In: ICML (2002)

\bibitem{gu2021domain}
Gu, Y., Tinn, R., Cheng, H., Lucas, M., Usuyama, N., Liu, X., Naumann, T., Gao, J., Poon, H.: Domain-specific language model pretraining for biomedical natural language processing. ACM Transactions on Computing for Healthcare (HEALTH)  \textbf{3}(1),  1--23 (2021)

\bibitem{hu2021lora}
Hu, E.J., Shen, Y., Wallis, P., Allen-Zhu, Z., Li, Y., Wang, S., Wang, L., Chen, W.: Lo{RA}: Low-rank adaptation of large language models. In: ICLR (2022)

\bibitem{irvin2019chexpert}
Irvin, J., Rajpurkar, P., Ko, M., Yu, Y., Ciurea-Ilcus, S., Chute, C., Marklund, H., Haghgoo, B., Ball, R., Shpanskaya, K., et~al.: Chexpert: A large chest radiograph dataset with uncertainty labels and expert comparison. In: Proceedings of the AAAI conference on artificial intelligence. vol.~33, pp. 590--597 (2019)

\bibitem{jiang2023mistral}
Jiang, A.Q., Sablayrolles, A., Mensch, A., Bamford, C., Chaplot, D.S., Casas, D.d.l., Bressand, F., Lengyel, G., Lample, G., Saulnier, L., et~al.: Mistral 7{B}. arXiv preprint arXiv:2310.06825  (2023)

\bibitem{lee2020biobert}
Lee, J., Yoon, W., Kim, S., Kim, D., Kim, S., So, C.H., Kang, J.: Bio{BERT}: a pre-trained biomedical language representation model for biomedical text mining. Bioinformatics  \textbf{36}(4),  1234--1240 (2020)

\bibitem{li2024llava}
Li, C., Wong, C., Zhang, S., Usuyama, N., Liu, H., Yang, J., Naumann, T., Poon, H., Gao, J.: Llava-med: Training a large language-and-vision assistant for biomedicine in one day (2023)

\bibitem{deepspine}
Lu, J.T., Pedemonte, S., Bizzo, B., Doyle, S., Andriole, K.P., Michalski, M.H., Gonzalez, R.G., Pomerantz, S.R.: Deep spine: Automated lumbar vertebral segmentation, disc-level designation, and spinal stenosis grading using deep learning. In: Machine Learning for Healthcare Conference. pp. 403--419. PMLR (2018)

\bibitem{luo2022biogpt}
Luo, R., Sun, L., Xia, Y., Qin, T., Zhang, S., Poon, H., Liu, T.Y.: Bio{GPT}: generative pre-trained transformer for biomedical text generation and mining. Briefings in Bioinformatics  \textbf{23}(6) (2022)

\bibitem{motohashi2024new}
Motohashi, M., Funauchi, Y., Adachi, T., Fujioka, T., Otaka, N., Kamiko, Y., Okada, T., Tateishi, U., Okawa, A., Yoshii, T., et~al.: A new deep learning algorithm for detecting spinal metastases on computed tomography images. Spine  \textbf{49}(6),  390--397 (2024)

\bibitem{nori2023can}
Nori, H., Lee, Y.T., Zhang, S., Carignan, D., Edgar, R., Fusi, N., King, N., Larson, J., Li, Y., Liu, W., et~al.: Can generalist foundation models outcompete special-purpose tuning? {C}ase study in medicine. Medicine  \textbf{84}(88.3),  77--3 (2023)

\bibitem{ouyang2022training}
Ouyang, L., Wu, J., Jiang, X., Almeida, D., Wainwright, C., Mishkin, P., Zhang, C., Agarwal, S., Slama, K., Ray, A., et~al.: Training language models to follow instructions with human feedback. In: NeurIPS (2022)

\bibitem{peng2018negbio}
Peng, Y., Wang, X., Lu, L., Bagheri, M., Summers, R., Lu, Z.: {N}eg{B}io: a high-performance tool for negation and uncertainty detection in radiology reports. AMIA Summits on Translational Science Proceedings  \textbf{2018}, ~188 (2018)

\bibitem{alpaca}
Taori, R., Gulrajani, I., Zhang, T., Dubois, Y., Li, X., Guestrin, C., Liang, P., Hashimoto, T.B.: Stanford {A}lpaca: An instruction-following {LLaMA} model. Tech. rep. (2023)

\bibitem{thirunavukarasu2023large}
Thirunavukarasu, A.J., Ting, D.S.J., Elangovan, K., Gutierrez, L., Tan, T.F., Ting, D.S.W.: Large language models in medicine. Nature medicine  \textbf{29}(8),  1930--1940 (2023)

\bibitem{touvron2023llama}
Touvron, H., Martin, L., Stone, K., Albert, P., Almahairi, A., Babaei, Y., Bashlykov, N., Batra, S., Bhargava, P., Bhosale, S., et~al.: Llama 2: Open foundation and fine-tuned chat models. Tech. rep. (2023)

\bibitem{Windsor22a}
Windsor, R., Jamaludin, A., Kadir, T., Zisserman, A.: Context-aware transformers for spinal cancer detection and radiological grading. In: International Conference on Medical Image Computing and Computer Assisted Intervention (2022)

\bibitem{windsor2024automated}
Windsor, R., Jamaludin, A., Kadir, T., Zisserman, A.: Automated detection, labelling and radiological grading of clinical spinal {MRI}s. Scientific Reports  \textbf{14}(1),  14993 (2024)

\bibitem{yan2022radbert}
Yan, A., McAuley, J., Lu, X., Du, J., Chang, E.Y., Gentili, A., Hsu, C.N.: Rad{BERT}: Adapting transformer-based language models to radiology. Radiology: Artificial Intelligence  \textbf{4}(4),  e210258 (2022)

\end{thebibliography}

\newpage
\appendix
\section{Appendix}

\begin{table}[]
\centering
\begin{tabular}{|p{1.4in}|p{3in}|}
\hline
\textbf{Condition} &
  \textbf{Definition} \\ \hline
\textbf{Cancer} &
  Spinal cancer includes malignant lesions that originate from the spine or spinal cord and metastatic or secondary tumours that have spread from another site to the spine. \\ \hline
\textbf{Stenosis} &
  Stenosis is any narrowing or compression of the spinal canal or nerves, including disc protrusions, impingement of nerve roots, or compromise of recesses. \\ \hline
\textbf{Spondylolisthesis} & Spondylolisthesis is a condition in which a vertebra slips out of place onto the bone below it. \\ \hline
\textbf{\Centerstack{Cauda Equina\\Compression}} & Cauda equina compression is the compression of a collection of nerve roots called the cauda equina, distinct from cauda equina syndrome; if the patient has cauda equina compression, the report will explicitly state its presence. \\ \hline
\textbf{Herniation} & Herniation is a condition in which a disc in the spine ruptures, and the disc nucleus is displaced from intervertebral space; it is more severe condition than disc protrusion or bulging, and if the patient has herniation, the report will explicitly state its presence. \\ \hline
\end{tabular}
\caption{\label{defs} Definitions used to prompt the label extraction pipeline.}
\end{table}

\begin{figure}
\centering
\includegraphics[width=\textwidth]{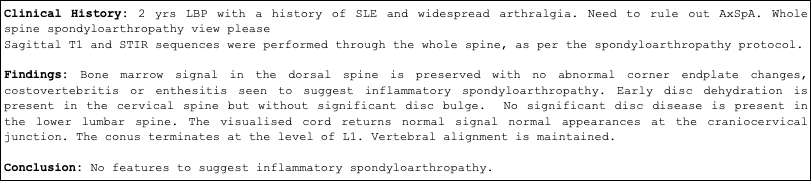}
\caption{\label{report_sections} An example report from \textbf{CancerData} with section headers shown in bold.} 
\end{figure}

\begin{figure}
\centering
\includegraphics[width=\linewidth]{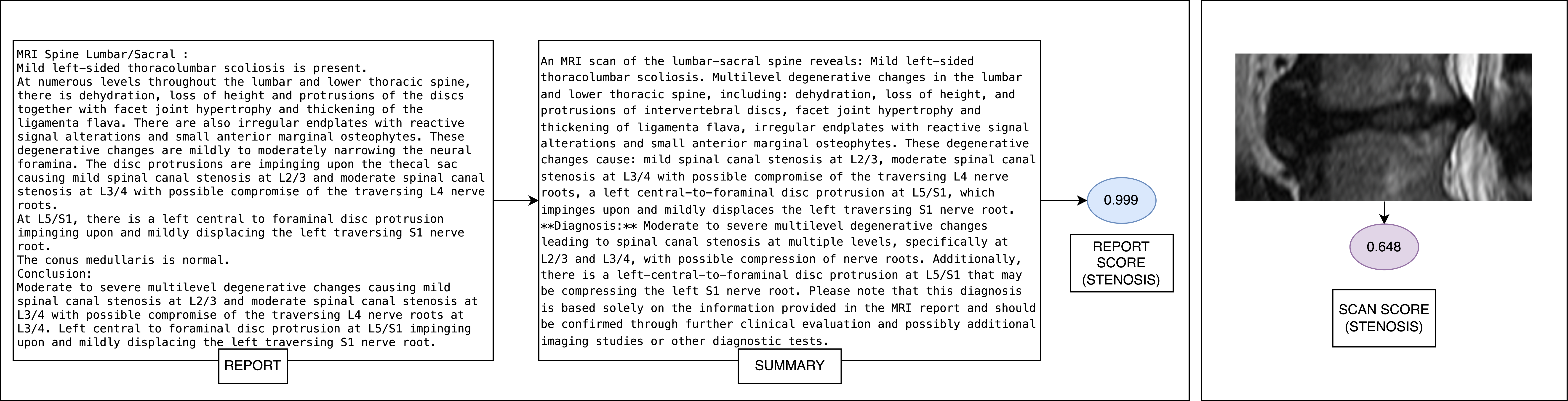}
\caption{\label{summ_result_sten} Example report and scan from \textbf{LumbarData} showing stenosis. Summary and scores were generated using our pipeline.} 
\end{figure}

\begin{figure}
\centering
\includegraphics[width=\linewidth]{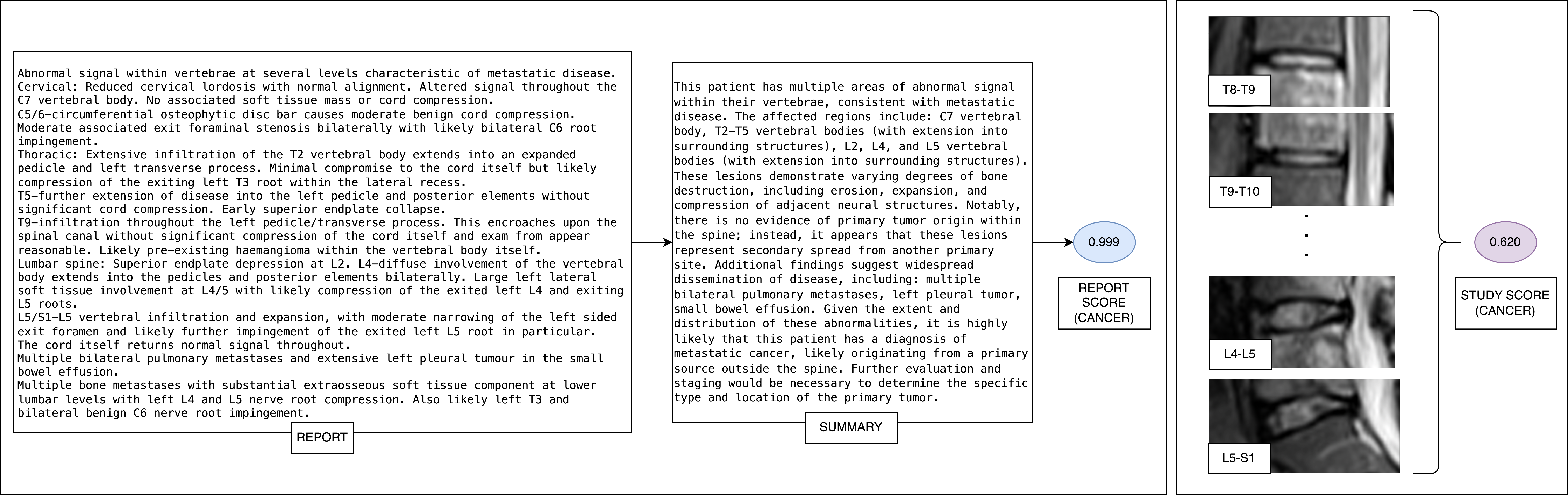}
\caption{\label{summ_result_cancer} Example report and scans from \textbf{CancerData} showing cancer. Summary and scores were generated using our pipeline.} 
\end{figure}

\begin{figure}
\centering
\includegraphics[width=\linewidth]{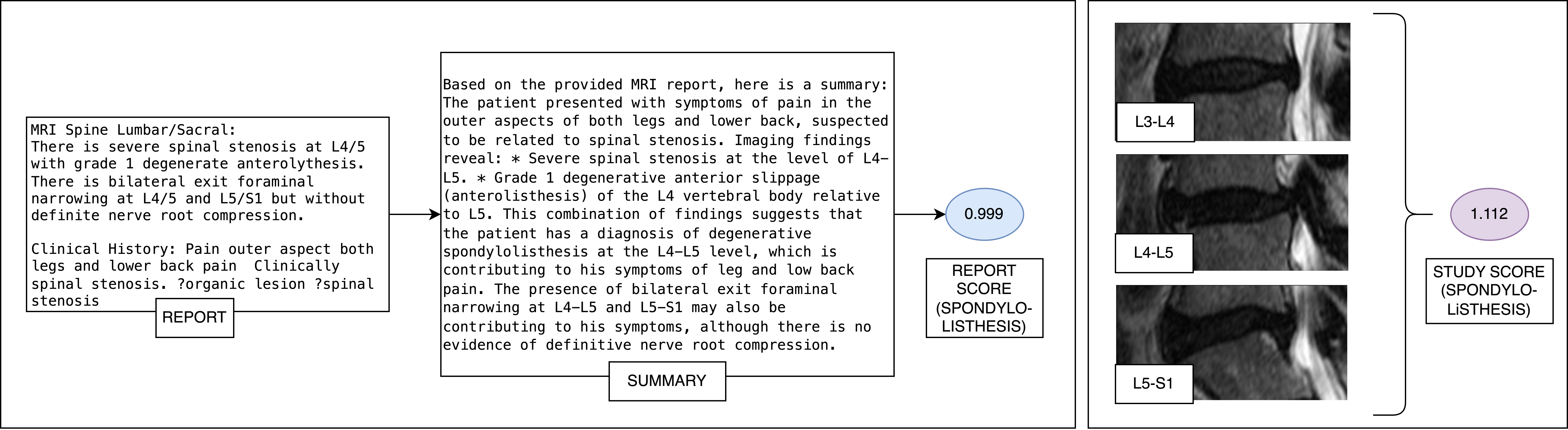}
\caption{\label{summ_result_spon} Example report and scans from \textbf{LumbarData} showing spondylolisthesis. Summary and scores were generated using our pipeline.} 
\end{figure}

\end{document}